\documentclass[aps,prb,twocolumn,floats,showpacs]{revtex4}
\usepackage{graphicx}
\usepackage{amsmath}
\def\stacksymbols #1#2#3#4{\def\theguybelow{#2}
    \def\verticalposition{\lower#3pt}
    \def\spacingwithinsymbol{\baselineskip0pt\lineskip#4pt}
    \mathrel{\mathpalette\intermediary#1}}
\def\intermediary#1#2{\verticalposition\vbox{\spacingwithinsymbol

      \everycr={}\tabskip0pt
      \halign{$\mathsurround0pt#1\hfil##\hfil$\crcr#2\crcr
               \theguybelow\crcr}}}

\begin{document}
\title{
Critical Level Statistics and Anomalously Localized States at the
Anderson Transition }

\author{H. Obuse and K. Yakubo}

\affiliation{Department of Applied Physics,
Graduate School of Engineering, Hokkaido University, Sapporo 060-8628, Japan.}

\begin{abstract}

We study the level-spacing distribution function $P(s)$ at the Anderson
transition by paying attention to anomalously localized states (ALS)
which contribute to statistical properties at the critical point. It is
found that the distribution $P(s)$ for level pairs of ALS coincides
with that for pairs of typical multifractal states. This implies that
ALS do not affect the shape of the critical level-spacing distribution
function. We also show that the insensitivity of $P(s)$ to ALS is a
consequence of multifractality in tail structures of ALS.

\end{abstract}

\pacs{71.30.+h, 73.20.Fz, 64.60.Ak, 71.70.Ej}

\maketitle

\section{Introduction}
\label{sec:1}

It is well recognized that a statistical description of spectral
correlations provides a powerful tool to study the disorder induced
metal-insulator transition, namely, the Anderson transition.
\cite{Gorkov1,Efetov1,Altshuler1,Altshuler2} Among several measures
representing spectral correlations, the nearest-neighbor level-spacing
distribution function $P(s)$ has been extensively studied so far, where
$s$ is the energy spacing between adjacent levels normalized by the
mean level spacing $\Delta$. The functional form of $P(s)$ is closely
related to the localization nature of corresponding wavefunctions.
According to the random matrix theory,\cite{Mehta1} the level-spacing
distribution function in the metallic phase is approximated by the
Wigner-Dyson distribution, namely,
\begin{equation}
P_{\text{W}}(s) = a_\beta s^\beta \exp(-c_\beta s^2),
\label{eq:1}
\end{equation}
where $\beta$ $( =1$, $2$, and $4$ for the orthogonal, the unitary, and
the symplectic ensembles, respectively) characterizes the universality
classes. The coefficients $a_\beta$ and $c_\beta$ are determined by the
normalization conditions $\int P_{\text{W}}(s) ds = \int s
P_{\text{W}}(s) ds =1$, as $a_1=\pi/2,$ $a_2=32/\pi^2,$
$a_4=2^{18}/3^6\pi^3,$ $c_1=\pi/4,$ $c_2=4/\pi,$ and $c_4=64/9\pi$. For
$s \ll 1$, the distribution function is proportional to $s^\beta$,
which implies that adjacent energy levels cannot approach each other
indefinitely because of mixing between two extended states. Since
$P_{\text{W}}(s) \propto s^4$ for $s \ll 1$ for the symplectic
ensemble, the level repulsion in this ensemble is strongest amongst the
three classes. In the insulating regime, mixing of quantum states
belonging to adjacent levels can be ignored and the energy levels are
uncorrelated. Consequently, the level-spacing distribution function
obeys the Poissonian,
\begin{equation}
P_{\text{P}}(s) = \exp(- s).
\label{eq:2}
\end{equation}
The Poisson distribution for the insulating phase and the three kinds
of Wigner-Dyson distribution functions for the metallic phase have been
confirmed numerically in two and three dimensions.
\cite{Evangelou1,Hofstetter1}

In addition to these two types of distribution functions, there exists
the third distribution at the critical point of the Anderson
transition.
\cite{Altshuler2,Shklovskii1,Ono1,Zharekeshev1,Varga1,Ndawana1} A
critical wavefunction percolates over the whole system in a
multifractal manner reflecting no characteristic length scales, and the
spatial profile of the wavefunction is similar to neither extended nor
localized states. \cite{Aoki1,Wegner1,Nakayama1} Since the level
statistics are governed by wavefunction profiles,\cite{Chalker1} the
critical level statistics are expected to be different from both
$P_{\text{W}}$ and $P_{\text{P}}$. In fact, it has been numerically
elucidated that the critical level-spacing distribution cannot be
expressed by either Eq.~(\ref{eq:1}) or (\ref{eq:2}) and is scale
independent and universal. The analytical form of the critical
distribution $P(s)$ is of great interest.

There exist attempts to solve this problem in a framework of the random
matrix theory. In the random matrix theory, a matrix ensemble is
characterized by the joint probability distribution of eigenvalues $\{
\varepsilon_i \}$. This probability distribution can be regarded as the
Gibbs function of one-dimensional interacting fictitious particles at
the inverse temperature $\beta(=1,2,4)$, whose positions are given by
$\varepsilon_i$. If these particles interact each other via a
logarithmic repulsion, namely, $V_{\text{int}}
(|\varepsilon_i-\varepsilon_j|)=-\ln |\varepsilon_i-\varepsilon_j|$ and
are in the confinement potential $V_{\text{pot}} (\varepsilon_i) =
\varepsilon_i^2$, the Gibbs function gives the Wigner-Dyson level
spacing distribution functions Eq.~(\ref{eq:1}). In order to obtain the
analytical form of $P(s)$ at criticality, two types of ingenuities have
been proposed. One is to set the power-law interaction $V_{\text{int}}
(|\varepsilon_i-\varepsilon_j|) \propto
|\varepsilon_i-\varepsilon_j|^{-\gamma}$ in the Gibbs function, where
$\gamma=1-1/d\nu$ and $\nu$ is the localization length
exponent.\cite{Aronov1,Kravtsov1} This leads the critical distribution
function proportional to $\exp(-c_{\beta}' s^{1+1/d \nu})$ for $s \gg
1$. The whole profile of $P(s)$ is then conjectured to be
\begin{equation}
P(s) = a_\beta' s^\beta \exp(-c_{\beta}' s^{1+1/d \nu}).
\label{eq:3}
\end{equation}
However, most of numerical results presented so far deviate from this
analytical
form.\cite{Schweitzer1,Evangelou2,Ohtsuki1,Kawarabayashi1,Zharekeshev2}
Although $P(s)$ for $s \ll 1$ is proportional to $s^\beta$, $P(s)$
obtained by numerical studies behaves as $\exp(-s/s_0)$ with $s_0 < 1$
in the large $s$ limit. Another attempt is to construct a random-matrix
ensemble with a log-squared potential $V_{\text{pot}} (\varepsilon_i) =
(\ln \varepsilon_i)^2/2a$.\cite{Kravtsov2,Nishigaki1} Using such a
non-Gaussian random-matrix ensemble, Nishigaki has obtained a
differential equation for the critical distribution
$P(s)$.\cite{Nishigaki1} Although the numerical solution of this
differential equation well reproduces the critical distribution $P(s)$
calculated by exact diagonalizations, the theory contains an ambiguous
parameter $a$ and the closed form of $P(s)$ remains to be unknown. Thus
the analytical form of $P(s)$ at the Anderson transition is still
unclear.

These arguments on $P(s)$ are based on the assumption that all of
critical states are multifractal characterized by an entire spectrum
comprising infinitely many exponents. Recently, it has, however, been
shown \cite{Obuse1} that anomalously localized states (ALS)
\cite{Altshuler3,Mirlin1,Muzykantskii1,Falko1,Uski1,Nikolic1,Kottos1}
in which most of amplitudes concentrate on a narrow spatial region
exist {\it with a finite probability in infinite systems} at the
Anderson transition point. ALS are brought by statistical fluctuations
in disorder realizations and show no multifractality because of the
characteristic length of large amplitude regions. It should be noted
that typical states are kept to be multifractal in critical systems.
When studying statistical properties of physical quantities defined at
the critical point, one should take notice of the influence of
ALS.\cite{Castillo1} In particular, ALS make a significant contribution
to distribution functions of critical quantities. If the level-spacing
distribution $P(s)$ calculated at the critical point is strongly
affected by ALS, there is a possibility that $P(s)$ for level pairs of
typical wavefunctions might be totally different from that obtained
numerically so far. It follows that the contribution of ALS should be
eliminated from a simulation result for checking numerically the
validity of an analytical expression of $P(s)$ at criticality. This is
because previous theoretical arguments are essentially based on the
random matrix theory which does not allow the existence of ALS.

In this paper, we study numerically the influence of ALS to the
level-spacing distribution function at the Anderson transition. Energy
levels and their corresponding wavefunctions at the critical point are
prepared in two-dimensional symplectic systems described by the SU(2)
model.\cite{Asada1} By employing our recently proposed definition of
ALS at criticality, we distinguish ALS from typical multifractal states
(MS) and construct ALS (MS) level-pair ensembles which contains only
pairs of energy levels corresponding to ALS (MS). It is found that the
critical level-spacing distribution function for an ALS level-pairs
ensemble coincides with that for MS level-pair ensembles and for the
entire ensemble. All characteristics in level-pair ensembles are
absorbed by their mean level spacings. This implies that the
elimination of ALS does not affect the profile of the level-spacing
distribution function. We also show that this remarkable feature of
$P(s)$ is a consequence of multifractality in tail structures of ALS.
This paper is organized as follows. In Sec.~\ref{sec:2}, we give a
quantitative definition of ALS at the critical point based on the idea
that ALS do not show multifractality. Level-pair ensembles constructed
for extracting contributions of ALS are also defined in this section.
Furthermore, a brief explanation of the SU(2) model employed in this
paper is given here. Calculated results are shown in Sec.~\ref{sec:3}.
An interpretation of our results and a condition that the critical
level-spacing distribution function should satisfy are presented in
Sec.~\ref{sec:4}, together with our conclusions.

\section{Definition of Anomalously Localized States and the model}
\label{sec:2}

In order to study the influence of ALS to the critical level-spacing
distribution function, it is necessary to distinguish ALS from a set of
critical wavefunctions. To this end, we employ an expediential
definition of ALS proposed in Ref.~18, which is based on the fact that
a typical critical wavefunction is multifractal while ALS are not.
Multifractal distributions of wavefunction amplitudes can be
characterized by the box-measure correlation function
$G_q(l,L,r)$,\cite{Janssen1}
\begin{equation}
G_q (l,L,r) = \frac{1}{N_b N_{b_r}} \sum_b \sum_{b_r} \mu_{b(l)}^q \mu_{b_{r}(l)}^q ,
\label{eq:4}
\end{equation}
where $\mu_{b(l)}=\sum_{i \in b(l)} |\psi_i|^2$ and
$\mu_{b_{r}(l)}=\sum_{i \in b_r(l)} |\psi_i|^2$ are the box measures of
squared amplitudes of a wavefunction $|\psi_i|^2$ in a box $b(l)$ of
size $l$ and in a box $b_r(l)$ of size $l$ fixed distance $r-l$ away
from the box $b(l)$, respectively, $N_b$ (or $N_{b_r}$) is the number
of boxes $b(l)$ [or $b_r(l)$], $q$ is the moment, and the summations
are taken over all such boxes in a system of size $L$. We concentrate
on the $l$ dependence of $G_2(l,L,r)$ for $r=l$ and the $r$ dependence
of $G_2(l,L,r)$ for $l=1$. Denoting $Q(l) \equiv G_2(l,L,r=l)$ and
$R(r) \equiv G_2(l=1,L,r)$, these two functions obey, if the
wavefunction is multifractal, the following power laws,\cite{Janssen1}
\begin{equation}
Q(l) \propto l^{d+\tau(4)},
\label{eq:5}
\end{equation}
and
\begin{equation}
R(r) \propto r^{-[d+2\tau(2)-\tau(4)]},
\label{eq:6}
\end{equation}
where $d$ and $\tau(q)$ are the spatial dimension and the mass exponent
for the $q$th moment of measures, respectively.\cite{Nakayama1}

We regard wavefunctions not satisfying Eqs.~(\ref{eq:5}) and
(\ref{eq:6}) as ALS at the critical point. To make this definition
quantitative, we introduce variances  $\text{Var} (\log_{10} Q)$ and
$\text{Var} (\log_{10} R)$ from functions $\log_{10} Q(l) = [d+\tau(4)]
\log_{10} l + c_Q$ and $\log_{10} R(r) = -[d+2
\tau(2)-\tau(4)]\log_{10} r + c_R$ obtained by the least-square fit for
a specific wavefunction. Non-multifractality of the wavefunction is
quantified by
\begin{equation}
\Gamma = \lambda \text{Var}(\log_{10} Q) + \text{Var} (\log_{10} R),
\label{eq:7}
\end{equation}
where $\lambda$ is a factor to compensate the difference between
average values of $\text{Var}(\log_{10} Q)$ and $\text{Var}(\log_{10}
R)$. The quantity $\Gamma$ is small (large) when a given wavefunction
is close to (far from) the typical multifractal state. We call $\Gamma$
the atypicality of the wavefunction. Thus one can expedientially regard
the wavefunctions with $\Gamma > \Gamma_{\text{ALS}}^*$ and $\Gamma <
\Gamma_{\text{MS}}^*$ as ALS and MS, respectively, where
$\Gamma_{\text{ALS}}^*$ and $\Gamma_{\text{MS}}^*$ are criterial values
of $\Gamma$ for ALS and MS.

Our aim is to calculate level-spacing distribution functions for ALS
and MS level-pair ensembles. The ALS level-pair ensemble
$S_{\text{ALS}}$ (the MS level-pair ensemble $S_{\text{MS}}$) is
defined as a set of pairs of adjacent levels whose corresponding
wavefunctions are both ALS (MS), namely,
\begin{equation}
S_{\text{ALS}}(\Gamma_{\text{ALS}}^*) =
\{ (\varepsilon, \varepsilon') : \Gamma > \Gamma_{\text{ALS}}^*  \ \text{and} \ \Gamma' > \Gamma_{\text{ALS}}^*\},
\label{eq:8}
\end{equation}
and
\begin{equation}
S_{\text{MS}}(\Gamma_{\text{MS}}^*) =
\{ (\varepsilon, \varepsilon') : \Gamma < \Gamma_{\text{MS}}^* \ \text{and} \ \Gamma' < \Gamma_{\text{MS}}^*\},
\label{eq:9}
\end{equation}
where $\varepsilon$ and $\varepsilon'$ are adjacent energy levels in
the critical region and $\Gamma$ and $\Gamma'$ are the atypicalities of
wavefunctions belonging to $\varepsilon$ and $\varepsilon'$,
respectively. In addition to $S_{\text{ALS}}$ and $S_{\text{MS}}$, the
ensemble of the whole pairs of adjacent levels in the critical region
is denoted by $S_0$, which is called the entire ensemble or the
original ensemble.

In the present paper, level-spacing distribution functions are
calculated for critical states in two-dimensional electron systems with
strong spin-orbit interactions (symplectic systems) because of the
advantage of system sizes. Among several models belonging to this
universality class, we adopt the SU(2) model, \cite{Asada1} because
scaling collections are known to be negligible due to a very short
spin-relaxation length. The Hamiltonian is compactly written in a
quaternion form as
\begin{equation}
\boldsymbol{H} = \sum_{i} \epsilon_i  \boldsymbol{c}_{i}^\dagger  \boldsymbol{c}_{i}
  -  V \sum_{i,j} \boldsymbol{R}_{ij} \boldsymbol{c}_{i}^\dagger \boldsymbol{c}_{j},
\label{eq:10}
\end{equation}
where $ \boldsymbol{c}_{i}^\dagger$ ($ \boldsymbol{c}_{i}$) is the
creation (annihilation) operator acting on a quaternion state
vector\cite{{Kyrala1}} and $\epsilon_i$ represents the on-site random
potential distributed uniformly in the interval $[-W/2, W/2]$.
(Quaternion-real quantities are denoted by bold symbols.) The strength
of the hopping $V$ is taken to be the unit of energy. The
quaternion-real hopping matrix element $\boldsymbol{R}_{ij}$ between
the sites $i$ and $j$ is given by
\begin{eqnarray}
\boldsymbol{R}_{ij} &=& \cos \alpha_{ij} \cos \beta_{ij} \boldsymbol{\tau}^0
+ \sin \gamma_{ij} \sin \beta_{ij}\boldsymbol{\tau}^1
\nonumber
\\
&-& \cos \gamma_{ij} \sin \beta_{ij}\boldsymbol{\tau}^2
+ \sin \alpha_{ij} \cos \beta_{ij}\boldsymbol{\tau}^3,
\label{eq:11}
\end{eqnarray}
for the nearest neighbor sites $i$ and $j$, and $\boldsymbol{R}_{ij}=0$
for otherwise. $\boldsymbol{\tau}^\mu (\mu=0,1,2,3)$ is the primitive
elements of the quaternion algebra. Random quantities $\alpha_{ij}$ and
$\gamma_{ij}$ are distributed uniformly in the range of $[0,2\pi)$, and
$\beta_{ij}$ is distributed according to the probability density
$P(\beta) d \beta = \sin(2\beta) d \beta$ for $0 \le \beta \le \pi/2$.
The critical disorder $W_c$ of this model is known to be $5.952$ for
the energy $E=1.0$. \cite{Asada1}

\section{Results}
\label{sec:3}

\begin{figure}[t]
\begin{center}
\includegraphics[width=8cm]{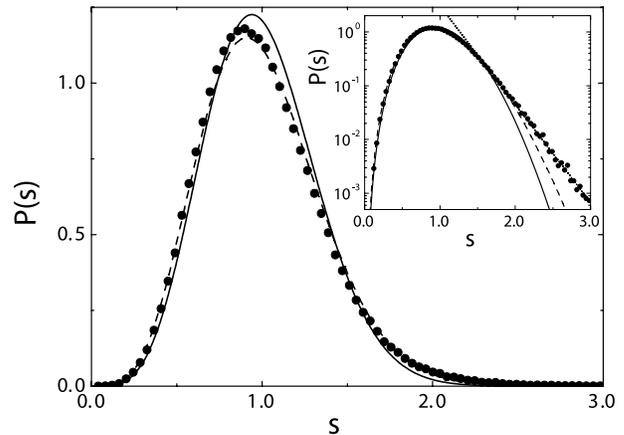}
\end{center}
\caption{Critical level-spacing distribution function for the entire
level-pair ensemble $S_0$ (circles). Solid and dashed lines represent
the Wigner-Dyson distribution function for the symplectic class
($\beta=4$) and the critical distribution predicted by Aronov {\it et
al.},\cite{Aronov1} namely, Eq.~(\ref{eq:3}) with $\nu=0.75$. The inset
shows the same plots in a semilogarithmic scale. Dotted line showing an
exponential decay is a guide to the eye. } \label{fig:1}
\end{figure}

We calculate the level-spacing distribution functions $P(s)$ for ALS
and MS level-pair ensembles at the metal-insulator transition point of
the two-dimensional SU(2) model. Periodic boundary conditions are
imposed in the both directions of systems of size $L=36$. The number of
disorder realizations is $2 \times 10^5$. We extract three successive
eigenenergies from one sample with $W=5.952$, which are closest to the
critical energy $E=1.0$. The total number of level spacings is then $4
\times 10^5$. All of these level spacings construct the entire
level-pair ensemble $S_0$. In contrast to conventional studies of the
level statistics, eigenstates are also required for distinguishing ALS
and MS from the set of critical states. The forced oscillator
method\cite{Nakayama2,Obuse2} for quaternion-real matrices has been
employed to calculate unfolded eigenvalues and the corresponding
eigenvectors.

\begin{table*}
\caption{List of the level-pair ensembles employed in the present work.
The ensemble $S_0$ is the entire (original) ensemble. The quantity
$\Gamma^*$ represents $\Gamma_{\text{ALS}}^*$ for ALS ensembles and
$\Gamma_{\text{MS}}^*$ for MS ones. $\Delta$ is the mean level spacing
of each ensemble, and $N$ is the number of level pairs included in the
ensemble.}\label{table:1}
\begin{ruledtabular}
\begin{tabular}{cccccccccc}
\multicolumn{1}{c}{}&
\multicolumn{1}{c}{}&
\multicolumn{1}{c}{\quad} &
\multicolumn{3}{c}{ALS Ensemble} &
\multicolumn{1}{c}{\quad} &
\multicolumn{3}{c}{MS Ensemble} \\ \hline
\multicolumn{1}{c}{Ensemble} &
\multicolumn{1}{c}{$S_0$} &
\multicolumn{1}{c}{\quad} &
\multicolumn{1}{c}{$S_{\text{ALS}}^1$} &
\multicolumn{1}{c}{$S_{\text{ALS}}^2$} &
\multicolumn{1}{c}{$S_{\text{ALS}}^3$} &
\multicolumn{1}{c}{\quad} &
\multicolumn{1}{c}{$S_{\text{MS}}^1$} &
\multicolumn{1}{c}{$S_{\text{MS}}^2$} &
\multicolumn{1}{c}{$S_{\text{MS}}^3$} \\
\multicolumn{1}{c}{$\Gamma^*$} &
\multicolumn{1}{c}{---} &
\multicolumn{1}{c}{\quad} &
\multicolumn{1}{c}{$0.03$} &
\multicolumn{1}{c}{$0.025$} &
\multicolumn{1}{c}{$0.01$} &
\multicolumn{1}{c}{\quad} &
\multicolumn{1}{c}{$0.01$} &
\multicolumn{1}{c}{$0.015$} &
\multicolumn{1}{c}{$0.03$} \\
$\Delta$ {$(\times 10^{-4})$} &
$8.12$ &
\multicolumn{1}{c}{\quad} &
$7.68$ & $7.74$ & $7.99$ & \quad &
$8.39$ & $8.33$ & $8.22$ \\
\multicolumn{1}{c}{$N$} &
\multicolumn{1}{c}{$400\,000$} &
\multicolumn{1}{c}{\quad} &
\multicolumn{1}{c}{$39\,100$} &
\multicolumn{1}{c}{$59\,154$} &
\multicolumn{1}{c}{$226\,802$} &
\multicolumn{1}{c}{\quad} &
\multicolumn{1}{c}{$32\,881$} &
\multicolumn{1}{c}{$82\,172$} &
\multicolumn{1}{c}{$227\,421$} \\
\end{tabular}
\end{ruledtabular}
\end{table*}

Figure \ref{fig:1} shows the level-spacing distribution $P(s)$ for the
entire level-pair ensemble $S_0$. The function $P(s)$ is proportional
to $s^4$ for $s\ll1$, while it is slightly shifted toward smaller $s$
as compared with the Wigner-Dyson distribution. In the limit of
$s\gg1$, $P(s)$ seems to decay exponentially as seen in the inset of
Fig.~\ref{fig:1}. These features of $P(s)$ are consistent with
numerical results reported previously.
\cite{Schweitzer1,Evangelou2,Ohtsuki1} In fact, when we try to fit our
data of $P(s)$ to the critical distribution function predicted by
Aronov {\it et al}.,\cite{Aronov1} the exponent $\nu$ in
Eq.~(\ref{eq:3}) is found to be $0.75$. This value is very close to
that obtained previously by a similar analysis for the Ando
model\cite{Ando1} as an alternative model of the two-dimensional
symplectic system,\cite{Ohtsuki1} though it largely deviates from the
value believed to be correct for this universality class ($\nu=2.8$).
This implies that previous numerical results with large energy windows
correctly represent the level-spacing distribution at criticality,
which differs from Eq.~(\ref{eq:3}) at large $s$ values but behaves
exponentially.

\begin{figure}[t]
\begin{center}
\includegraphics[width=8cm]{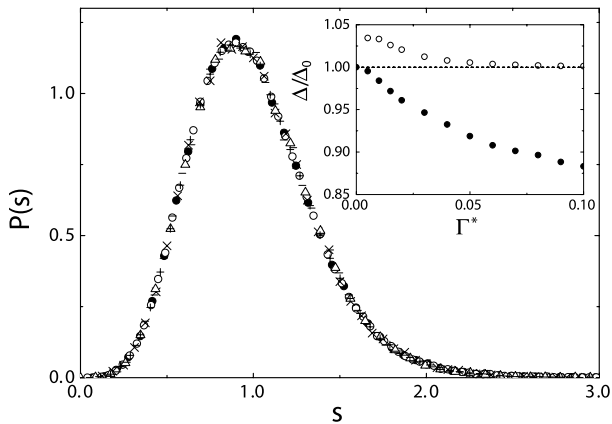}
\end{center}
\caption{Critical level-spacing distribution functions for ALS and MS
level-pair ensembles. Symbols $\triangle$, $-$, $\times$, $+$, and
$\circ$ represent $P(s)$ for level-pair ensembles $S_{\text{ALS}}^1$,
$S_{\text{ALS}}^2$, $S_{\text{MS}}^1$, $S_{\text{MS}}^2$, and $S_0$,
respectively. Filled circles show $P(s)$ for the mixed level-pair
ensemble $S_{\text{mix}}$ with $\Gamma^*=0.01$. The inset exhibits the
mean level spacings for ALS level-pair ensembles (filled circles) and
MS ensembles (open circles) rescaled by the mean spacing $\Delta_0$ for
$S_0$. The abscissa indicates $\Gamma_{\text{ALS}}^*$ for filled
circles and $\Gamma_{\text{MS}}^*$ for open circles. } \label{fig:2}
\end{figure}

We have prepared several level-pair ensembles $S_{\text{ALS}}$ and
$S_{\text{MS}}$ by choosing values of $\Gamma_{\text{ALS}}^*$ and
$\Gamma_{\text{MS}}^*$ to calculate the level-spacing distributions for
these ensembles. The parameter $\lambda$ in the definition of $\Gamma$
[see Eq.~(\ref{eq:7})] is set to be $3.0$, because the average of
$\text{Var}(\log_{10} R)$ is about three times larger than that of
$\text{Var}(\log_{10} Q)$, while the choice of $\lambda$ does not
affect our conclusions. Statistical data of seven ensembles used in the
present work is listed in Table \ref{table:1}.

Figure \ref{fig:2} shows the level-spacing distribution functions for
the ensembles of $S_{\text{ALS}}^1$, $S_{\text{ALS}}^2$,
$S_{\text{MS}}^1$, $S_{\text{MS}}^2$, and $S_0$ by different symbols.
We should remark that the distribution functions are normalized and the
spacing $s$ is rescaled by the mean level spacing $\Delta$ for each
ensemble listed in Table \ref{table:1}. The distribution function for
$S_0$ is the same with that shown in Fig.~\ref{fig:1}. It is surprising
that all distributions collapse onto a single universal curve which is
identical to the conventional critical level-spacing distribution
function for the two-dimensional symplectic class. This seems to
conflict with the relation between the spectral correlation and
wavefunction profiles. Naively, the function $P(s)$ for an ALS
level-pair ensemble $S_{\text{ALS}}$ would be more Poisson-like due to
the localized nature of wavefunctions. Our numerical result indicates
that ALS at the critical point do not affect the critical level-spacing
distribution at all in spite of the fact that the probability to find
ALS at criticality is finite even in an infinite system. This implies
that $P(s)$'s obtained by previous numerical works without paying
attention to ALS represent the spectral correlation for typical
critical states and can be used for judging the validity of theories
for the critical level-spacing distribution based on the random matrix
theory. In the next section, we will discuss the reason of this
remarkable property of $P(s)$ at criticality. The inset of
Fig.~\ref{fig:2} shows mean level spacings for ALS and MS level-pair
ensembles rescaled by the mean level spacing $\Delta_0$ for the entire
level-pair ensemble $S_0$. The criteria $\Gamma^*$ of level-pair
ensembles means $\Gamma_{\text{ALS}}^*$ for ALS level-pair ensembles
and $\Gamma_{\text{MS}}^*$ for MS ensembles. In the case of ALS
level-pair ensembles (filled circles), the mean level spacing decreases
from $\Delta_0$ with increasing $\Gamma_{\text{ALS}}^*$, while for MS
level-pair ensembles (open circles) it increases as decreasing
$\Gamma_{\text{MS}}^*$ . This implies that ALS level pairs mainly
contribute to the level-spacing distribution in a small $s$ region as
demonstrated by Table \ref{table:1}, which is consistent with the fact
that the repulsive force between two adjacent levels corresponding to
localized states is weak.

\begin{figure*}[t]
\begin{center}
\includegraphics[width=17.5cm]{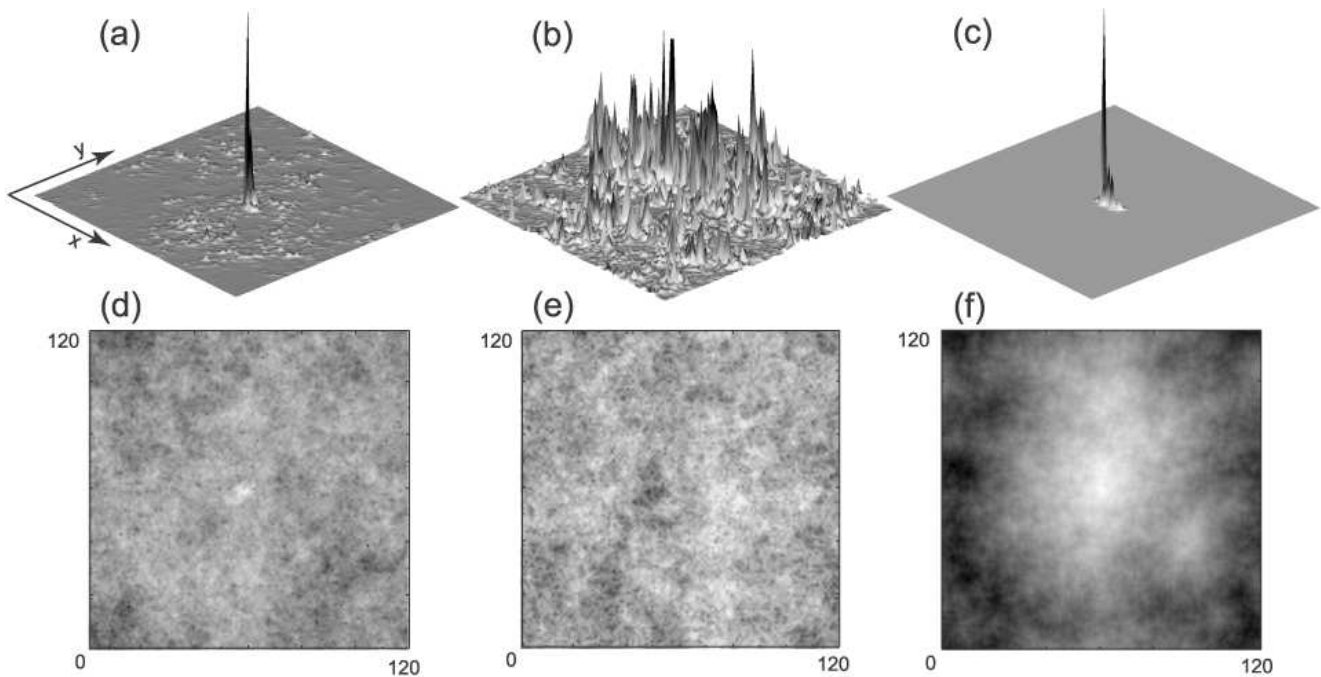}
\end{center}
\caption{Spatial profiles of an ALS [(a) and (d)], a MS [(b) and (e)],
and a truly localized [(c) and (f)] wavefunctions. These wavefunctions
are computed for the SU(2) model of size $L=120$. The truly localized
wavefunction [(c) and (f)] is an eigenstate of the system in the
insulating phase ($W=15.0$ and $E=1.0$). Figures (d)-(f) represent the
logarithm of squared amplitudes by gray-scale plots. } \label{fig:3}
\end{figure*}

\section{Discussions and Conclusions}
\label{sec:4}

Let us consider the reason why the functional form of the critical
level-spacing distribution $P(s)$ is not affected by ALS as shown in
the previous section. To answer this question, it is important to
examine carefully the spatial profile of ALS. Figure \ref{fig:3} shows
squared amplitude distributions of an ALS ($\Gamma=0.100$) at
criticality, a typical MS ($\Gamma=0.009$), and a localized state in
the insulating phase ($W=15.0$ and $E=1.0$) of the SU(2) model of size
$L=120$. We see from the upper row of Fig.~\ref{fig:3} that amplitudes
of the ALS wavefunction [Fig.~\ref{fig:3}(a)] concentrate on a narrow
region in the system, which resembles the usual localized state
[Fig.~\ref{fig:3}(c)] in appearance and is in contrast to the MS
wavefunction [Fig.~\ref{fig:3}(b)]. There exists, however, a crucial
difference between profiles of ALS and truly localized states. The
difference can be found in {\it tails} of wavefunctions. The lower row
of Fig.~\ref{fig:3} shows the gray-scale plots of the logarithm of
squared amplitudes corresponding to the right above wavefunctions. For
the truly localized state [Fig.~\ref{fig:3}(f)], amplitudes in
logarithmic scale decrease with getting away from the localization
center (the center of the gray-scale plot), which means that the
wavefunction decays exponentially. On the contrary, the ALS
wavefunction does not possess such an exponential tail as demonstrated
clearly in Fig.~\ref{fig:3}(d). It should be noted that the gray-scale
range of Fig.~\ref{fig:3}(f) ($10^{-36}$ to $10^{-1}$) is much wider
than that of Fig.~\ref{fig:3}(d) ($10^{-10}$ to $10^{-2}$). A
remarkable feature of the ALS wavefunction is that the amplitude
distribution away from the peak position of the ALS [the center of
Fig.~\ref{fig:3}(d)] is quite similar to that shown in
Fig.~\ref{fig:3}(e). This implies that amplitudes in the tail region of
ALS distribute in a multifractal manner.

\begin{figure}[t]
\begin{center}
\includegraphics[width=8cm]{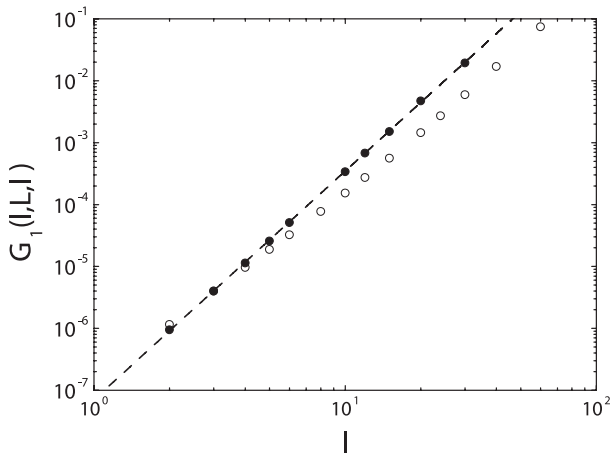}
\end{center}
\caption{Box-measure correlation functions $G_q(l,L,r)$ for $q=1$ and
$r=l$ for the whole eigenstate depicted by Fig.~\ref{fig:3}(a) (open
circles, $L=120$) and the part of the wavefunction containing a tail
region cut from the original ALS wavefunction(filled circles,
$L=L_{\text{s}}=60$). Dashed line shows the least square fit for
$G_1(l,L_{\text{s}},l)$ for the extracted wavefunction.} \label{fig:4}
\end{figure}

To confirm this perspective, we performed the multifractal analysis for
the amplitude distribution only in the tail region of ALS. For this
purpose, we have extracted a part of the eigenstate depicted by
Fig.~\ref{fig:3}(a) within a $60 \times 60$ subsystem cut from the
original system ($120 \times 120$) so that the extracted eigenstate
contains only a tail region of the whole eigenstate. (For cutting the
subsystem, the original eigenstate has been appropriately shifted by
taking the periodic boundary conditions into account.) In order to
analyze its multifractal properties, we have renormalized the extracted
wavefunction and calculated the box-measure correlation function
$G_q(l,L,r)$ for $q=1$, $L=L_{\text{s}}$, and $r=l$, where
$L_{\text{s}}$ is the size of the subsystem ($L_{\text{s}}=60$). Figure
\ref{fig:4} shows that the calculated $G_1(l,L_{\text{s}},l)$ is
proportional to $l^{d+D_2}$ with $D_2=1.69\pm0.01$, where $D_2=\tau(2)$
is the correlation dimension, while the correlation function
$G_1(l,L,l)$ for the whole wavefunction is not. This value of $D_2$ is
very close to the value reported so far for typical multifractal
wavefunctions belonging to this universality class.\cite{Obuse1} This
implies that tail regions of ALS exhibit the same multifractality with
that of typical critical wavefunctions.\cite{Falko2,Apalkov1}

The tail structure of ALS gives a clue to understanding the behavior of
$P(s)$ at criticality for the ALS level-pair ensemble $S_{\text{ALS}}$.
Due to non-vanishing multifractal tails of ALS, quantum states
belonging to adjacent ALS levels weakly couple each other through the
overlap between wavefunctions in their tail region. As a consequence,
we expect a repulsive force between these two ALS levels. The strength
of the repulsive force is weaker than that for MS level pairs, because
the average amplitude in the ALS tail region is small compared to the
average of MS amplitudes. This explains a small mean level spacing for
$S_{\text{ALS}}$ as presented in Table \ref{table:1} and the inset of
Fig.~\ref{fig:2}. Since the spectral correlation is governed by
overlaps of multifractal tails of ALS wavefunctions, it is reasonable
that the critical level-spacing distribution for an ALS level-pair
ensemble has the same functional form with $P(s)$ for MS level-pair
ensembles. Multifractality in tails of ALS does not depend on the value
of $\Gamma_{\text{ALS}}^*$. Therefore, the critical level-spacing
distributions for the entire level-pair ensemble $S_0$, the ALS
ensemble $S_{\text{ALS}}$ with any $\Gamma_{\text{ALS}}^*$, and the MS
ensemble $S_{\text{MS}}$ with any $\Gamma_{\text{MS}}^*$ have the same
profile, and all characteristics of level-spacing ensembles are
absorbed by the mean level spacings.

\begin{figure}[t]
\begin{center}
\includegraphics[width=8cm]{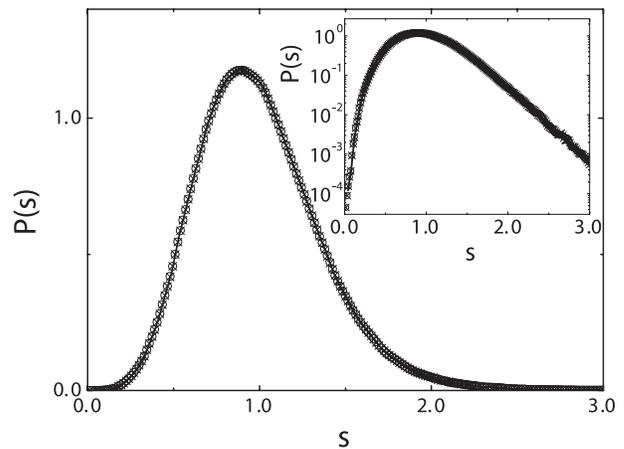}
\end{center}
\caption{Numerical confirmation of Eq.~(\ref{eq:15}). Solid line
represents the critical level-spacing distribution function calculated
for the entire level-pair ensemble $S_0$. Open circles and crosses are
obtained from the right-hand side of Eq.~(\ref{eq:15}) with
$\Gamma^*=0.01$ and $0.03$, respectively. The inset shows the same
plots in a semilogarithmic scale. } \label{fig:5}
\end{figure}

The fact that the critical level-spacing distribution is invariant
under changes of level-pair ensembles gives a condition for $P(s)$. If
the ALS criteria $\Gamma_{\text{ALS}}^*$ is equal to
$\Gamma_{\text{MS}}^*$ (say $\Gamma^*$), the entire level-pair ensemble
$S_0$ can be decomposed into three ensembles $S_{\text{ALS}}$,
$S_{\text{MS}}$ and $S_{\text{mix}}$, where $S_{\text{mix}}$ contains
level pairs of ALS and MS, namely, $S_{\text{mix}}(\Gamma^*) = \{
(\varepsilon, \varepsilon') : \Gamma > \Gamma^* \ \text{and} \ \Gamma'
< \Gamma^*\}$. These ensembles have no overlaps each other, i.e.,
$S_{\text{ALS}} \cap S_{\text{MS}} = S_{\text{MS}} \cap S_{\text{mix}}
= S_{\text{mix}} \cap S_{\text{ALS}} = \phi$ (null set) and
$S_0=S_{\text{ALS}} \cup S_{\text{MS}} \cup S_{\text{mix}}$. In
addition to $P(s)$'s for $S_{\text{ALS}}$ and $S_{\text{MS}}$, we have
also confirmed that the distribution $P(s)$ for $S_{\text{mix}}$ has
the same functional form with $P(s)$ for $S_0$, $S_{\text{ALS}}$, or
$S_{\text{MS}}$ (see filled circles in Fig.~\ref{fig:2}). Thus, we have
\begin{equation}
P(s)=P_{\text{ALS}}(s)=P_{\text{MS}}(s)=P_{\text{mix}}(s),
\label{eq:12}
\end{equation}
for any value of $\Gamma^*$. Here, $P_x(s)$ is the critical
distribution for $S_x$ ($x$ stands for the suffix $\lq\lq \text{ALS}
"$, $\lq\lq \text{MS} "$, or $\lq\lq \text{mix} "$). For
$\Gamma_{\text{ALS}}^*=\Gamma_{\text{MS}}^*$, $P(s)$ can be expressed
as a sum of contributions from $S_{\text{ALS}}$, $S_{\text{MS}}$, and
$S_{\text{mix}}$:
\begin{equation}
P(s)=\tilde{P}_{\text{ALS}}(s)+\tilde{P}_{\text{MS}}(s)+\tilde{P}_{\text{mix}}(s),
\label{eq:13}
\end{equation}
where the distribution $\tilde{P}_x(s)$ is defined to be a function of
the level spacing rescaled by the mean level spacing $\Delta_0$ for
$S_0$ (not $S_x$) and normalized as $\int \tilde{P}_x(s) ds = N_x/N_0$,
where $N_0$ and $N_x$ are the numbers of level pairs included in the
ensembles $S_0$ and $S_x$, respectively. [$P_x(s)$ is a function of the
spacing rescaled by $\Delta_x$ for $S_x$ and normalized to unity.]
Taking into account differences in the normalization conditions and the
meanings of $s$, $\tilde{P}_x(s)$ is related to $P_x(s)$ as
\begin{equation}
\tilde{P}_x(s) = \frac{n_x}{\delta_x} P_x(s/\delta_x),
\label{eq:14}
\end{equation}
where $n_x=N_x/N_0$ and $\delta_x=\Delta_x/\Delta_0$. Using
Eqs.~(\ref{eq:12})-(\ref{eq:14}), we obtain
\begin{eqnarray}
P(s)&=&\frac{n_{\text{ALS}}}{\delta_{\text{ALS}}} P(s/\delta_{\text{ALS}}) \nonumber \\
&+&\frac{n_{\text{MS}}}{\delta_{\text{MS}}} P(s/\delta_{\text{MS}})
+\frac{n_{\text{mix}}}{\delta_{\text{mix}}} P(s/\delta_{\text{mix}}).
\label{eq:15}
\end{eqnarray}
Note that quantities $n_x$ and $\delta_x$ depend on $\Gamma^*$. The
critical level-spacing distribution function $P(s)$ should satisfy the
above condition for any value of $\Gamma^*$.\cite{com1} Figure \ref{fig:5} shows
numerically the validity of this condition. Data fabricated by using
the right-hand side of Eq.~(\ref{eq:15}) for $\Gamma^*=0.01$ and $0.03$
(symbols) agree quite well with the critical distribution itself (solid
line).

We should note here that Eq.~(\ref{eq:15}) is not consistent with the
exponential behavior of the critical level-spacing distribution
function in the large $s$ limit, which is numerically observed in the
inset of Fig.~\ref{fig:1}. Thus, Eq.~(\ref{eq:12}) [or
Eq.~(\ref{eq:15})] implies a {\it possibility} that the profile of
$P(s)$ for $s \gg 1$ is not exactly exponential. For clarifying the
asymptotic behavior of $P(s)$ for $s \gg 1$ or the validity of
Eq.~(\ref{eq:12}), further investigations are required.

In conclusion, we have studied the level-spacing distribution function
$P(s)$ at the Anderson transition point of the two-dimensional SU(2)
model belonging to the symplectic class by paying attention to ALS.
Since ALS at criticality exist with a finite probability even in the
thermodynamic limit, quantities at the critical point might be
influenced by ALS. Facts that ALS seem to have very different
wavefunction profiles from typical MS ones and $P(s)$ is governed by
spatial structures of wavefunctions intimate that the level-spacing
distribution is greatly affected by ALS. To examine the influence of
ALS to $P(s)$ at the critical point, we prepared numerically several
ALS (MS) level-pair ensembles which are constructed by pairs of ALS
(MS) levels. Our remarkable result shows that level-spacing
distributions for ALS level-pair ensembles coincide completely with
those for MS and the entire ensembles, while the mean level spacing
$\Delta$ for the ALS ensemble becomes smaller than $\Delta$ for the MS
ensemble. Our findings imply that the spectral correlation for typical
critical states can be evaluated even without eliminating ALS pairs
from the original level-pair ensemble. Since ALS do not influence the
function $P(s)$, the critical level-spacing distribution function can
be understood by an analytical argument based on the random matrix
theory in which ALS are not considered. We have also shown that the
property of $P(s)$ insensitive to the existence of ALS is a consequence
of multifractality in ALS tail structures. Furthermore, it has been
pointed out that $P(s)$ should satisfy the condition derived from the
invariance of $P(s)$ under changes of level-pair ensembles. Although
these results were obtained for the two-dimensional SU(2) model, we
believe that the significant property of $P(s)$ is common in other
universality classes.

\begin{acknowledgements}
We are grateful to T. Nakayama for helpful discussions. This work was
supported in part by a Grant-in-Aid for Scientific Research from Japan
Society for the Promotion of Science (No.~$14540317$). Numerical
calculations in this work have been mainly performed on the facilities
of the Supercomputer Center, Institute for Solid State Physics,
University of Tokyo.
\end{acknowledgements}


\begin{thebibliography}{99}
\bibitem{Obuse2}
Details of the forced oscillator method for quaternion-real matrices
will be published elsewhere.
\bibitem{Falko2}
The power law behavior in a ALS tail region has been predicted by
Ref.~30 in a metallic phase of two-dimensional symplectic systems.
\bibitem{com1}
Three terms in the right-hand side of Eq.~(\ref{eq:15}) are single-peak
functions with different peak positions. This does not mean that the
sum of these three terms has three peaks, because the peak positions are very
close each other.
\end{thebibliography}
\end{document}